\documentclass[preprint]{aastex}
\usepackage{graphicx}
\usepackage{color}

\newcommand{\fermi}{{\it Fermi} LAT}
\newcommand{\hess}{{H.E.S.S.}}
\newcommand{\veritas}{{VERITAS}}
\newcommand{\nustar}{{\it NuSTAR}}

\begin{document}
\title{Origin of X-ray and gamma-ray emission from the Galactic central region}
\shorttitle{X-rays and gamma-rays from the Galactic center}
\shortauthors{Chernyshov et al.}

\author{D. O. Chernyshov\altaffilmark{1,2,3}, K.-S. Cheng\altaffilmark{2}, V. A. Dogiel\altaffilmark{1,2,4}, C. M.~Ko\altaffilmark{3,5}}

\altaffiltext{1}{I.E.Tamm Theoretical Physics Division of P.N.Lebedev Institute of Physics, Leninskii pr. 53, 119991 Moscow, Russia;
\texttt{chernyshov@lpi.ru}}
\altaffiltext{2}{Department of Physics, University of Hong Kong, Pokfulam Road, Hong Kong, China}
\altaffiltext{3}{Institute of Astronomy, National Central University, Jhongli, Taoyuan, Taiwan, R.O.C.}
\altaffiltext{4}{Moscow Institute of Physics and Technology, 141700 Moscow Region, Dolgoprudnii, Russia}
\altaffiltext{5}{Department of Physics and Center for Complex Systems, National Central University,
Jhongli, Taoyuan, Taiwan, R.O.C.;
\texttt{cmko@astro.ncu.edu.tw}}

\begin{abstract}
We study a possible connection between different non-thermal emissions from the inner few parsecs of the Galaxy.
We analyze the origin of the gamma-ray source 2FGL J1745.6$-$2858 (or 3FGL J1745.6$-$2859c) in the Galactic Center (GC) and
the diffuse hard X-ray component recently found by {\it NuSTAR}, as well as the radio emission and processes of hydrogen ionization
from this area.
We assume that a source in the GC injected energetic particles with power-law spectrum into the surrounding medium in the past
or continues to inject until now. The energetic particles may be protons, electrons or a combination of both.
These particles diffuse to the surrounding medium and interact with gas, magnetic field and background photons to produce
non-thermal emissions.
We study the spectral and spatial features of the hard X-ray emission and gamma-ray emission by the particles from the central source.
Our goal is to examine whether the hard X-ray and gamma-ray emissions have a common origin.

Our estimations show that in the case of pure hadronic models the expected flux of hard X-ray emission is too low.
Despite protons can produce a non-zero contribution in gamma-ray emission, it is unlikely that they and their secondary electrons
can make a significant contribution in hard X-ray flux.
In the case of pure leptonic models it is possible to reproduce both X-ray and gamma-ray emissions
for both transient and continuous supply models.
However, in the case of continuous supply model the ionization rate of molecular hydrogen may significantly exceed the observed value.
\end{abstract}


\date{\today}

\maketitle

\section{Introduction}\label{sec:intro}
The interstellar medium of several pc around Sgr A$^*$ in the center of our Galaxy is characterized by a number of peculiar parameters
\citep[see, e.g.,][]{katia07,katia12}.
The central supermassive black hole is surrounded by a circumnuclear disk (CND) whose total mass was estimated by \citet{chr05}
to be $10^6$ $M_\odot$.
The analysis of \citet{katia12} gave a slightly lower value of mass about $2\times 10^5$ $M_\odot$ for the region of radius
$R_c=3\sim 5$ pc with average gas density in the CND of about $4\times 10^5$ cm$^{-3}$.
Unlike other regions of the Galaxy the central region shows prominent emissions in a very broad range of electromagnetic waves,
from radio to gamma-ray.

\begin{itemize}
\item
The Very Energetic Radiation Imaging Telescope Array System (\veritas) \citep{veritas16}
and The High Energy Stereoscopic System (\hess) \citep{aha09} detected a prominent gamma-ray flux in the TeV energy range.
Recent observations of \hess~ \citep{hess-16} with the angular resolution $0.01\degr$ provided some new information about this GC area.
These observations found a point-like source in the center and a diffuse emission around the source that is correlated with
the gas distribution which may mean a hadronic origin of this emission. The
recovered cosmic ray density
decreases as $1/r$
(where $r$ is the distance from the source) that can be interpreted as a stationary ejection of relativistic protons by the source.

\item
The Large Area Telescope (\fermi) of {\it Fermi} also detected a gamma-ray source in the GeV region \citep[see][]{nolan12,3fgl}.
In the second \fermi~ source catalog, this source was identified as the source 2FGL J1745.6$-$2858.
The estimated gamma-ray flux from this source for $E>2$ GeV is $I_{\rm obs}=1.08\times 10^{-10}$ erg cm$^{-2}$ s$^{-1}$
with a spectral index $\gamma=2.68$ \citep[see][]{masha11}. This corresponds to a luminosity about $8\times 10^{35}$ erg s$^{-1}$.
The positional error circle of 2FGL J1745.6$-$2858 overlaps with Sgr A$^*$. This is compatible with the picture that
its emission originates within the CND or within the central cavity.
It is reasonable to assume that this emission is provided either by CRs protons \citep{masha11, linden12},
or by CR electrons \citep{kusunose, masha15}, or by both protons and electrons \citep{guo13}
produced in the vicinity of Sgr A$^*$.

Independent analysis of the \fermi~ data by \citet{masha11} and \citet{masha15} indicated that
recovered by them spectra are compatible with 2FGL J1745.6$-$2858 data at GeV energies yet significantly softer in the sub-GeV range.

In the \fermi~ 4-Year Point Source Catalog (3FGL) \citep{3fgl}, this source was identified as 3FGL J1745.6$-$2859c.
However the gamma-ray spectrum of this source is significantly softer below 1 GeV in comparison to 2FGL J1745.6$-$2858.
The discrepancy may be due to another bright source 3FGL J1745.3$-$2903c in this area, which was not identified in the second catalog.
Since there is a flag ``c'' in the name of the source the reliability of the derived spectrum of 3FGL J1745.6$-$2859c is not very high
due to the contamination from the nearby bright source.
In this paper we do not judge which analysis is more appropriate, instead we present our model fittings
for both 3FGL data and data obtained by \citet{masha15}.

\item
Large scale diffuse radio emission from this region is known as radio halo. It has a spherical shape with radius of about 8 pc.
At frequencies about tens of GHz the emission have a clear synchrotron nature and thus confirms the presence of high-energy electrons
in this area \citep{pedlar89}.

\item
The Nuclear Spectroscopic Telescope Array (\nustar) found a flux of non-thermal hard X-ray emission in the direction of
the central few parsecs region of the Galaxy \citep{perez15}. Total luminosity in the energy range 20-40 keV was estimated
as $2.4\times 10^{34}$ erg s$^{-1}$, which corresponds to an energy flux at the Earth of $3.3\times 10^{-12}$ erg cm$^{-2}$ s$^{-1}$.
 Spatial profile of the emission has an elliptical shape with a major axis 8 pc and a minor axis 4 pc. It is larger than gas
structures in the GC like the central cavity (1.2 pc in diameter) and CND (6 pc in diameter) \citep{katia12}. It also does not spatially
coincide with Sgr A East. However it looks similar to the Nuclear Star Cluster
\citep[8 pc by 5.6 pc,][]{schodel}.
Therefore it was concluded that this emission was produced by unresolved point like sources.
This idea was further developed by \citet{hailey16} who concluded that
the X-ray emission can be produced by intermediate polars (a type of cataclysmic variables) with masses of about 0.9 solar masses.

Despite the fact that point-like sources potentially play a major role in the hard X-ray emission we suppose that there is still room for a diffuse
component of X-rays generated by CRs. Indeed as we mentioned earlier the presence of high-energy protons or electrons is confirmed by gamma-ray
and radio observations. It is perceivable that these energetic particles also contribute to the total hard X-ray flux from the GC.

In principle, hard X-ray photons may be generated by synchrotron emission of secondary high-energy electrons
produced by collisions of TeV protons with the background gas in the GC region.
Recent observations of TeV gamma-ray point-like source detected a clear cut-off in the gamma-ray spectrum \citep{veritas16} at around 12 TeV.
This gives the maximum energy of primary protons at about $12/0.075=160$ TeV. The maximum energy of secondary electrons
is thus about $12\times 0.039 / 0.075=6.24$ TeV \citep[see][]{atoyan}.
Therefore, in order to produce a non-thermal emission with power-law spectrum up to 40 keV by secondary electrons,
the magnetic field strength there should be about 16 mG.
This value is significantly higher than the $1\sim 3$ mG obtained by \citet{kill92, yusef96, eat13}.
Hence we must conclude that synchrotron origin of this X-rays is doubtful.

If however the cut-off in the spectrum of this source is due to absorption of the gamma-ray photons by their interactions
with background infra-red emission as suggested by \citet{hess-16}, the spectrum of primary protons may be power-law
up to  energies higher than 140 TeV. In this case synchrotron emission can produce up to 10\%-20\% of the flux detected by
\nustar~ depending on the magnetic field strength. We note however that the density of the infra-red photons does not appear
to be high enough to produce the observed cut-off in the gamma-ray spectrum by
photon-photon ($\gamma\gamma$) collisions \citep[see, e.g.,][]{kist15}.

Alternatively, X-rays can be generated by electron bremsstrahlung or inverse Compton effect (or both).
This model was developed in \citet{chern14}. They assumed that the gamma-ray flux of the source 2FGL J1745.6$-$2858
is generated by relativistic protons in CND. The secondary electrons produce a flux of hard X-rays by bremsstrahlung in CND
and by inverse Compton effect in the region surrounding CND. Then the X-ray halo around CND could be more extended
than the gas distribution as observed by \nustar. However, \citet{chern14} considered a spherically symmetric model
which is unable to explain the observed asymmetry of the hard X-ray emission.
Besides, the hard X-ray flux predicted in that work turned out to be significantly lower than the observed value.
Therefore, modification of the model is required if this interpretation is correct.
\end{itemize}

Thus, these observations indicate a high efficiency CR production in the GC.
Additional evidence supporting high density of CRs in the GC came from the measured ionization rate in the 1 pc region of CND by
\citet{goto13,goto14}.
They obtained a value $\zeta \simeq 1.2\times 10^{-15}$ s$^{-1}$, which is one order of magnitude higher than in other parts of the Galaxy.
This ionization is most likely produced by subrelativistic CRs with a density higher than those outside the GC region.
The estimated source luminosity of these subrelativistic CRs in the GC is about
$10^{38}\sim 10^{39}$ erg s$^{-1}$ \citep[see, e.g.,][]{dog13,dog14,yus3}.

The goal of our present investigation is to explain simultaneously all phenomena mentioned above
except TeV gamma-ray emission
in the framework of a single model where CRs are injected from the central source.
This source may be stationary \citep{macias,hess-16} or transient \citep{macias}.
Both possibilities will be discussed. We will estimate the required parameters of this source.

We analyze two types of models of the source:
\begin{itemize}
\item
{\it Hadronic model}. The central source injects mainly high energy protons.
Most of the non-thermal electrons is produced by interactions of these protons with background medium.
Gamma-rays are generated by
proton-proton (pp) collisions.
X-rays are produced by protons via inverse bremsstrahlung and by secondary electrons through bremsstrahlung and inverse Compton scattering.
Radio emission is produced by synchrotron losses of secondary electrons.
\item
{\it Leptonic model}. The central source injects primary high energy electrons.
In this case gamma-ray and hard X-ray emission are mainly produced by their bremsstrahlung and
inverse Compton scattering. Radio flux is produced by their synchrotron emission.
\end{itemize}

Unlike previous investigations our goal is to explain all types of non-thermal emissions
using the same injection of particles (protons or electrons).
The ultimate goal is to reproduce the observed hard X-ray spectrum and the spatial distribution of hard X-rays
using the observed gamma-ray spectrum as a reference point to determine the injection and propagation parameters.
Also we examine whether the estimated gamma-ray flux in the models is compatible with other observations there,
such as the radio emission and the ionization rate of molecular hydrogen.
Below we take into account the non-spherical shape of the gas distribution in the central few parsecs.
The parameters of the gas distribution in the CND region are taken from \citet{katia12}.

In this investigation we assume that TeV gamma-ray emission discovered by \hess~ and \veritas, and GeV gamma-ray emission discovered by
\fermi~ are produced by two different sources. Indeed the overall gamma-ray spectrum indicates an ``ankle-like'' break at the transition from GeV to TeV
energies. This break can be described naturally by a combination of two components with different spectral indices. Since
TeV component exhibits much harder spectrum, its contribution to X-rays, radio emission and ionization should be much lower in comparison to
GeV component. Therefore we restrict our analysis to GeV gamma-rays only.

\section{Model description}\label{sec:model}
We use a cylindrically symmetric model to study the evolution of proton and electron distribution functions
$f_p(t,r,z,E)$ and $f_e(t,r,z,E)$ with time.
The $z = 0$ plane corresponds to the CND plane of symmetry, and the origin $(r,z) = (0,0)$
corresponds to the center of the CND with Sgr A$^*$ as the source. Equation for the distribution functions is
\begin{eqnarray}\label{eq:kin_eq_p}
\frac{\partial f_{p,e}}{\partial t} - \frac{\partial }{\partial z}\left(D_{p,e}\frac{\partial f_{p,e}}{\partial z}\right)
- \frac{1}{r}\frac{\partial }{\partial r}\left(rD_{p,e}\frac{\partial f_{p,e}}{\partial r}\right)
+ \frac{\partial }{\partial E}\left[\left(\frac{dE}{dt}\right)_{p,e} f_{p,e}\right] = Q_{p,e}(t,r,z,E)\,.
\end{eqnarray}
Here we assume that the diffusion is isotropic and the diffusion coefficient $D_{p,e} = D_{p,e}(E,r,z)$ has different values
inside and outside the molecular cloud. The diffusion coefficient is a function of particle rigidity,

\begin{equation}\label{eq:diffusion_gen}
D_{p,e}(E,r,z) = D_0(r,z)\beta\left(\frac{p}{p_0}\right)^\varsigma \,,
\end{equation}
where the value of $D_0$ is the same for protons and electrons, $p$ is the particle momentum,
$\beta$ is the particle velocity in the units of $c$,
$p_0 = 4$ GeV$/c$, and the spectral index $\varsigma$ is determined by the spectrum of MHD-turbulence
for non-magnetized particles or by a structure of magnetic field lines for magnetized particles.
Outside the cloud we assume that the diffusion is mainly due to energetic particle scattering by the interstellar
turbulence with Kolmogorov spectrum and therefore $\varsigma^{\rm inter} = 0.33$ \citep{acker12}.

Inside the molecular cloud, however, the situation is quite different due to the very efficient damping of MHD waves
by ion-neutral friction.
In this case diffusion is due to tangled magnetic field. Therefore the mean-free path of charged particles is determined
by the characteristic correlation length of the intracloud magnetic field. As a result the diffusion coefficient
does not depend on particle rigidity and $\varsigma^{\rm cloud} = 0$ \citep[see, e.g.,][]{dog15}.

The parameter $D_0$ in the GC center region and inside the CND is unknown.
For some regions of the Galaxy, it can be estimated from observational data.
\citet{acker12} gave a value of $D^{\rm inter}_0 = 5\sim 10\times 10^{28}$ cm$^2$ s$^{-1}$
as an average over the Galaxy diffusion coefficient.
As for intracloud, diffusion estimations for cloud such as Sgr B2 gave a value of about
$D^{\rm cloud}_0 = 10^{28}$ cm$^2$ s$^{-1}$ \citep{dog15}.

However, the inner few parsec of the Galaxy are located in a special region characterized by processes of
effective energy release in different forms. Therefore, it is reasonable to assume that the diffusion coefficient there is
significantly smaller than in the other part of the GC region.
We estimate this parameter from the spatial and spectral properties of the emissions.

The energy loss term $dE/dt$ in Equation~(\ref{eq:kin_eq_p}) consist of Coulomb losses and proton-proton losses for protons
\citep{mns94}, and of Coulomb, bremsstrahlung, inverse-Compton and synchrotron losses for electrons \citep{blu70}.
The source term $Q_{p,e}(t,r,z,E)$ are model-dependent and we derive them later (see sections~\ref{sec:hadronic} and \ref{sec:leptonic}).

\begin{figure}[h]
\begin{center}
\includegraphics[width=0.8\textwidth]{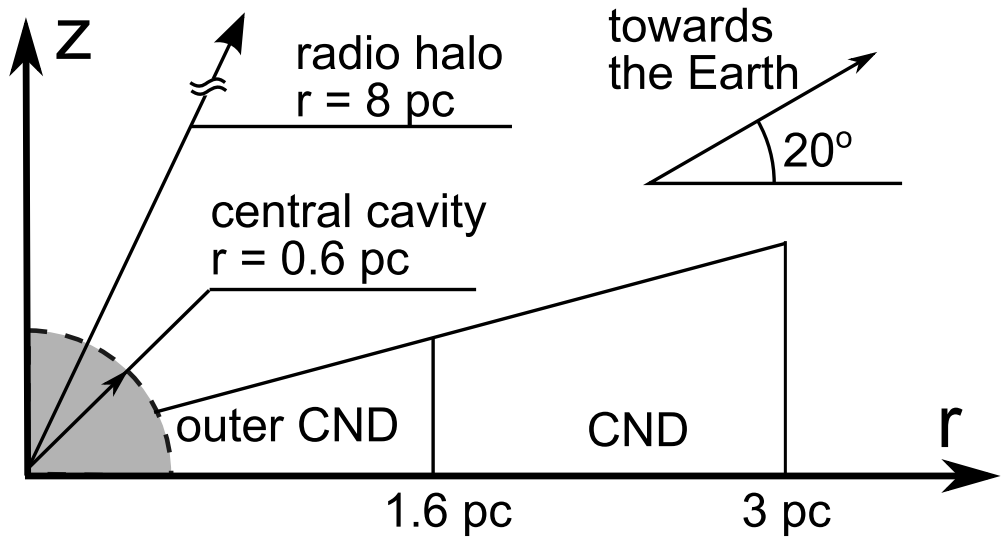}
\end{center}
\caption{Gaseous structure in the Galactic Center region \citep{katia12}.}
\label{fig:gasmap}
\end{figure}

We adopt the ambient density distribution from \citet{katia12}. We divide the space into the following regions:
(1) central cavity with density of $10^3$ cm$^{-3}$, (2) outer layer of CND with density of $3.2\times 10^4$ cm$^{-3}$,
(3) CND with density of $4.4\times 10^5$ cm$^{-3}$, and (4) radio halo with density of $210$ cm$^{-3}$ (see Figure~\ref{fig:gasmap}).

In order to estimate the electron energy loss and their emission we need to know the distribution of soft photons.
We take into account the following components (assuming they all have black-body spectrum):
(1) optical emission from the central cluster ($T=3\times 10^4$ K), (2) mid-infrared emission from the central cluster ($T= 170$ K),
(3) dust emission from CND ($T = 70$ K), and (4) optical emission from the nuclear star cluster ($T = 3500$ K).
Details on the estimation of soft photon density can be found in Appendix~\ref{appx:wop}.

From equation~(\ref{eq:kin_eq_p}) we derive the evolution of the distribution functions $f_p$ and $f_e$ and the gamma-ray emission produced by these CRs
\begin{eqnarray}\label{eq:eq_gammarays}
I_\gamma(t,E_\gamma) = \int dV&\left[\int dE ~f_p(t,r,z,E) n(r,z)v \left(\frac{d\sigma(E,E_\gamma)}{dE_\gamma}\right)_{pp}\right.  \nonumber\\
&+\int dE ~f_e(t,r,z,E) w(r,z) v \left(\frac{d\sigma(E,E_\gamma)}{dE_\gamma}\right)_{\rm IC}  \nonumber \\
&+\int dE ~f_e(t,r,z,E) n(r,z) v \left. \left(\frac{d\sigma(E,E_\gamma)}{dE_\gamma}\right)_{\rm br} \right] \,,
\end{eqnarray}
where $v$ is particles velocity, $n(r,z)$ and $w(r,z)$ are the density distributions of ambient gas and soft photons, respectively.
Proton-proton cross-section $(d\sigma/dE_\gamma)_{pp}$ is taken from \citet{kamae}, bremsstrahlung $(d\sigma/dE_\gamma)_{\rm br}$
and inverse-Compton cross-sections $(d\sigma/dE_\gamma)_{\rm IC}$ are taken from \citet{blu70}.

We adjust the initial parameters to make sure that gamma-ray spectrum $I_\gamma(E_\gamma)$ matches the observed one by \fermi.
With these parameters X-ray emissivity can be estimated as
\begin{eqnarray}\label{eq:eq_xrays}
\epsilon(t,r,z,E_x) = &\int dE ~f_p(t,r,z,E) n(r,z)v \left(\frac{d\sigma(E,E_x)}{dE_x}\right)_{\rm ib}  \nonumber\\
&+\int dE ~f_e(t,r,z,E) w(r,z) v \left(\frac{d\sigma(E,E_x)}{dE_x}\right)_{\rm IC} \nonumber \\
&+\int dE ~f_e(t,r,z,E) n(r,z) v \left(\frac{d\sigma(E,E_x)}{dE_x}\right)_{\rm br} \,,
\end{eqnarray}
where $(d\sigma(E,E_x)/dE_x)_{\rm ib}$ is the cross-section of the inverse-bremsstrahlung process \citep{haya,tati03}.

To obtain the spatial distribution of X-rays across the sky, the emissivity should be integrated along the line-of-sight.
We take into account the fact that CND is inclined to the Galactic plane by $20\degr$ (see Figure~\ref{fig:gasmap}).

Ionization rate of molecular hydrogen is computed from
\begin{equation}\label{eq:eq_ioniz}
\zeta = \int dE ~\sigma_{Hp} vf_p + \int dE ~\sigma_{He} vf_e\,,
\end{equation}
where $\sigma_{Hp}$ and $\sigma_{He}$ are the cross-sections of hydrogen ionization by proton and electron impact, respectively
\citep{tati03}.
We combine Equation~(\ref{eq:eq_ioniz}) with the approach of \citet{dalg} to take into account of the influence low-energy electrons.

Equations~(\ref{eq:eq_gammarays}), (\ref{eq:eq_xrays}) and (\ref{eq:eq_ioniz}) are applicable to both hadronic and leptonic models.
Obviously, in the case of leptonic model we assume $f_p \equiv 0$ and ignore all terms related to protons.

\section{Hadronic models}\label{sec:hadronic}
We start from injection spectrum of primary  protons.
Energetic electrons are considered as secondaries and their source function is
\begin{equation}\label{eq:Qesec}
Q_e(t,r,z,E) = n(r,z)\int dE_p f_p(t,r,z,E_p) v \left(\frac{d\sigma(E_p,E)}{dE}\right)_{\rm se}\,,
\end{equation}
where the cross-section $(d\sigma/dE)_{\rm se}$ of electron production includes
proton-proton collision term \citep{kamae} and the knock-on term \citep{haya}.

The source function of protons is model-dependent. We adopt the following form,
\begin{equation}\label{eq:eq_Q_genform}
Q_p(t,r,z,E) = A(E)\,T(t)\,\delta(z)\,\frac{\delta(r)}{2\pi r}\,,
\end{equation}
where $\delta(z)$ and $\delta(r)$ are Dirac delta-functions of $z$ and $r$, respectively.
$A(E)$ is the spectrum of the injected particles and $T(t)$ describes the temporal variations of the injection.

The injected particle spectrum is assumed to be a power-law in momentum space, i.e.,
\begin{equation}\label{eq:eq_A_momentum}
\frac{dN(>p)}{dp} \propto p^{-\alpha}\,,
\end{equation}
where $p$ is the momentum of the particle, $\alpha$ is a spectral index of of the injection spectrum
and $N(>p)$ is total number of particles with momenta higher than $p$.
One should note that a momentum power-law distribution
of particles is expected inside of accelerator while the injected spectrum of particles may be modified by energy losses or escape
which will potentially develop a spectral break.
This is really important since we perform calculations in a wide energy range - from non-relativistic to
ultra-relativistic energies. However in the particular case of the hadronic model we would like to maximize potential X-ray emission so
we assume that particles are injected with power-law spectrum without any breaks.

After re-expressing Equation~(\ref{eq:eq_A_momentum}) in terms of kinetic energy $E$,
we obtain the spectrum $A(E)$ in Equation~(\ref{eq:eq_Q_genform})
\begin{equation}\label{eq:q_pwlaw}
A(E)=A_0\,({E+Mc^2})\,{(E^2+2Mc^2E)^{-(\alpha+1)/2}} \,,
\end{equation}
where $E$ and $M$ are the kinetic energy and mass of the particle, respectively.
In the present section we take $M = m_p$, the mass of proton.

The normalization constant $A_0$ and the spectral index $\alpha$ of Equation~(\ref{eq:q_pwlaw}) can be estimated from the gamma-ray data.
Their values should be adjusted to fit the observed gamma-ray spectrum.
With these parameters one can estimate the intensity of hard X-rays and radio emission.

Radio emission depends strongly on the magnetic field strength whose value has not been well established.
Hence radio emission cannot be considered as a strong indicator for the validity of the model.
The same is true for the ionization rate whose value is known only up to an order of magnitude.
Therefore, we consider that these two observations play a supplementary role in our discussion, and
we mainly concentrate on hard X-ray emission to examine whether the model is compatible with observations or not.

\subsection{Single-burst hadronic model}\label{sec:sbhm}
In the framework of this model, we assume that the injection happened some time ago in the form of a short burst.
The temporal term of the injection function in Equation~(\ref{eq:eq_Q_genform}) is expressed as $T(t)=\delta(t)$.

\begin{figure}[h]
\begin{center}
\includegraphics[width=0.9\textwidth]{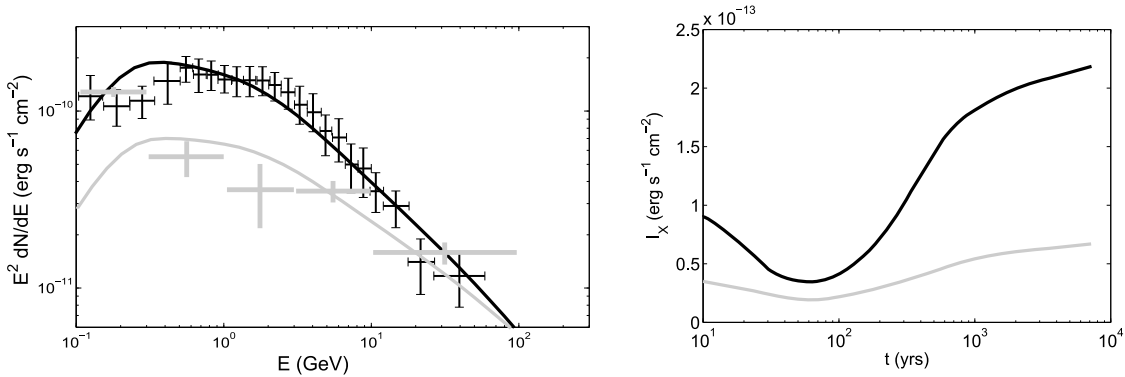}
\end{center}
\caption{{\it Left:} Gamma-ray spectrum in the case of single-burst hadronic model.
Black crosses with tick mark are data points taken from \citet{masha15} and gray crosses from 3FGL \citep{3fgl}.
{\it Right:} Evolution of the hard X-ray intensity in the case of single-burst hadronic model.
In both panels, theoretical curves corresponding to \citet{masha15} and 3FGL \citep{3fgl} are black and gray, respectively.
}
\label{fig:hadronic}
\end{figure}

The observed spectrum of \citet{masha15} can be reproduced by assuming $\alpha = 2.5$ in Equation~(\ref{eq:q_pwlaw}),
and $\alpha = 2.3$ for the 3FGL data \citep{3fgl} (see the left panel of Figure~\ref{fig:hadronic}).
As one can see from the figure the fit of the data is not very good. It was indicated by \citet{masha15} that
the low-energy part of the spectrum cannot be explained in the framework of pure hadronic model, and the situation
is worse in the case of 3FGL data.

Gamma-ray spectrum depends weakly on time $t$ and the diffusion coefficients.
However the intensity of hard X-ray emission is more sensitive to these parameters.
Its value increases as the diffusion coefficients decrease.
To maximize the intensity of X-rays we used the following propagation parameters:
inside the cloud $D^{\rm cloud}_0 = 10^{26}$ cm$^2$ s$^{-1}$ while outside the cloud $D^{\rm inter}_0 = 10^{27}$ cm$^2$ s$^{-1}$
(see Equation~\ref{eq:diffusion_gen}).
The evolution of the hard X-ray intensity with time is presented in the right panel of Figure~\ref{fig:hadronic}.

\begin{figure}[h]
\begin{center}
\includegraphics[width=0.9\textwidth]{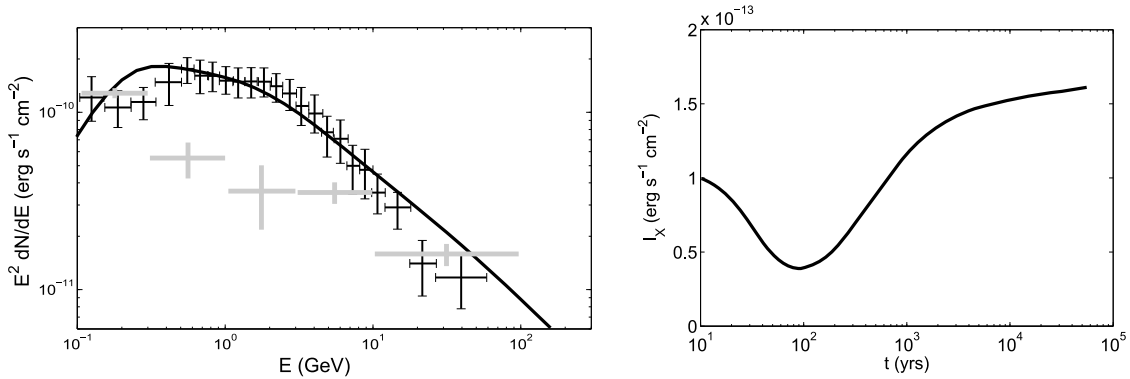}
\end{center}
\caption{The same as Figure~\ref{fig:hadronic} except for the case of continuous supply hadronic model.
}
\label{fig:hadronic_stat}
\end{figure}

The resulting X-ray intensity in the model is significantly less than the value of $3.3\times 10^{-12}$ erg cm$^{-2}$ s$^{-2}$
observed by \nustar. This conclusion agrees with \citet{chern14}.
Although the curves in right panel of Figure~\ref{fig:hadronic} show a tendency to increase with time,
an unreasonably long time and unreasonably high energy of the initial burst are needed to reach the observed value.
Therefore, we conclude that single-burst hadronic model is unable to reproduce the observed hard X-ray emission from the GC.
However, relativistic protons may produce a non-negligible contribution to the gamma-ray emission.

The total energy budget to produce gamma-ray emission depends on how long ago the injection of protons took place. Since protons
escape from the region with dense gas, the longer the time from the injection, the higher the energy is required. For characteristic time
about $2000\sim 7000$ yrs, the total energy required is within $(0.15\sim 1) \times 10^{50}$ erg. One can see that a single supernova
explosion can supply enough energy assuming 10\% acceleration efficiency.

\subsection{Continuous supply hadronic model}\label{sec:cshm}
In this case the temporal term of the injection function in Equation~(\ref{eq:eq_Q_genform}) is expressed as
$T(t)=\Theta(t)$, where $\Theta(t)$ is the Heaviside step function.

The results are shown in Figure~\ref{fig:hadronic_stat}.
The total power required is about $5\times 10^{37}$ erg/s.
Both gamma-ray and X-ray results do not differ strongly from that of the single-burst hadronic model except that
the X-ray emission is even lower.
Therefore, the conclusion is not in favor of continuous supply (or stationary) model also.

\section{Leptonic models}\label{sec:leptonic}
In this section we assume that the central source injects relativistic electrons.
The source function of electrons in Equation~(\ref{eq:kin_eq_p}) is similar to that used in Section~\ref{sec:hadronic}
(see Equation~(\ref{eq:eq_Q_genform})),
\begin{equation}\label{eq:eq_Q_genform_e}
Q_e(t,r,z,E) = A(E)\,\Theta(E_{\rm max}-E)\,T(t)\,\delta(z)\,\frac{\delta(r)}{2\pi r}\,,
\end{equation}
where the energy part $A(E)$ is described by Equation~(\ref{eq:q_pwlaw}) with $M = m_e$, the electron mass.
We introduced a cut-off energy $E_{\rm max}$ (the Heaviside step-function in Equation~(\ref{eq:eq_Q_genform_e}))
to reproduce a turn-over in the gamma-ray spectrum at 2 GeV.
We note that this cut-off can occur when the electrons pass through a region with very high magnetic or radiation field
before they escape into the interstellar medium \citep{masha15}.

Similar to hadronic scenario we would like to maximize potential X-ray emission,
we assume that particles are injected with power-law spectrum without any breaks. However it is necessary to mention that spectrum of electrons
below 100 MeV is heavily influenced by ionization losses and therefore is weakly sensitive to the spectral index of the injected spectrum.
We estimate the potential impact of a break in the injected spectrum of electrons below.

Taking all these changes into account we repeat the same procedures in Section~\ref{sec:hadronic}.

\subsection{Single-burst leptonic model}\label{sec:sblm}
Similar to Section~\ref{sec:sbhm},
the temporal term of the injection function in Equation~(\ref{eq:eq_Q_genform_e}) is expressed as $T(t)=\delta(t)$.

For the observed gamma-ray spectrum of \citep{masha15}, in order to reproduce the cut-off in the data at about 2 GeV,
we take two sets of value of spectral index and cut-off energy, (i) $\alpha = 2.1$ with $E_{\rm max} = 50$ GeV,
and (ii) $\alpha = 2.3$ with $E_{\rm max} = 100$ GeV.
For 3FGL data, we take  $\alpha = 2.5$, and choose $E_{\rm max}\rightarrow\infty$
(or simply neglect the Heaviside step-function in Equation~(\ref{eq:eq_Q_genform_e}))
because there is no spectral break in 3FGL data and the high energy cut-off is not necessary.

In the case of leptonic models, X-ray and gamma-ray emission consists of bremsstrahlung and inverse-Compton components.
The energy losses through inverse-Compton effect produces a softer spectrum in X-rays than by bremsstrahlung.
Therefore, for the same intensity of gamma-ray emission one would expect higher intensity of X-rays
if inverse-Compton effect dominates.

Thus, in order to maximize the intensity of X-rays (with a given gamma-ray intensity),
it is necessary to reduce the bremsstrahlung component of the emission as much as possible.
To achieve this effect, we adjust values of the diffusion coefficients.
Similar to the case of hadronic model we choose fairly low value of diffusion coefficients inside the cloud
$D^{\rm cloud}_0 = 10^{26}$ cm$^2$ s$^{-1}$ (see Equation~\ref{eq:diffusion_gen}).
Outside the cloud, however, we use a much larger value of the diffusion coefficient $D^{\rm inter}_0 = 10^{28}$ cm$^2$ s$^{-1}$
to prevent the particles propagating back to the cloud.

Spatial distribution of hard X-ray emission is time-dependent. To reproduce the observed spatial profile of X-rays with
the assumed diffusion coefficients, it is necessary to allow at least $7\times 10^2$ years has passed after the injection.

\begin{figure}[h]
\begin{center}
\includegraphics[width=0.9\textwidth]{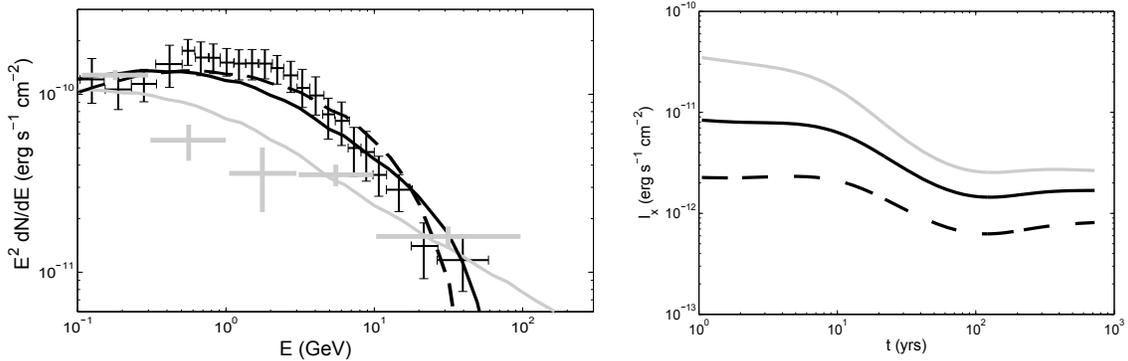}
\end{center}
\caption{The same as Figure~\ref{fig:hadronic} except for the case of single-burst leptonic models. Black solid lines correspond to
$\alpha = 2.3$ and dashed lines to $\alpha = 2.1$. Gray lines are based on 3FGL data with $\alpha = 2.5$.
}
\label{fig:leptonic}
\end{figure}

Gamma-ray spectra produced in leptonic models are shown in the left panel of Figure~\ref{fig:leptonic}.
The black dashed line is the case of $\alpha = 2.1$ and black solid line is the case of $\alpha = 2.3$.
The gray line corresponds to $\alpha = 2.5$ in 3FGL data.
We present the solution at $t = 7\times 10^2$ years.
One can see that the gamma-ray emission is reproduced quite well in the case of $\delta = 2.1$ but for the case of $\delta = 2.3$
there is a deficit around a few GeV.

The evolution of the intensity of hard X-rays is shown in the right panel of Figure~\ref{fig:leptonic}.
Once again, the black dashed line is the case of $\alpha = 2.1$ and black solid line is the case of $\alpha = 2.3$,
and the gray line corresponds to $\alpha = 2.5$ in 3FGL data.
One can see that if the spectrum of electrons is harder (smaller $\alpha$), the intensity of hard X-ray emission is lower.
In all cases, the intensity of hard X-rays (at $t\sim 10^3$ years) is within a factor of two around the value
$3.3\times 10^{-12}$ erg cm$^2$ s$^{-1}$ observed by \nustar.
Since the uncertainty of the X-ray flux is high, this can be considered as a good fit.

\begin{figure}[h]
\begin{center}
\includegraphics[width=0.9\textwidth]{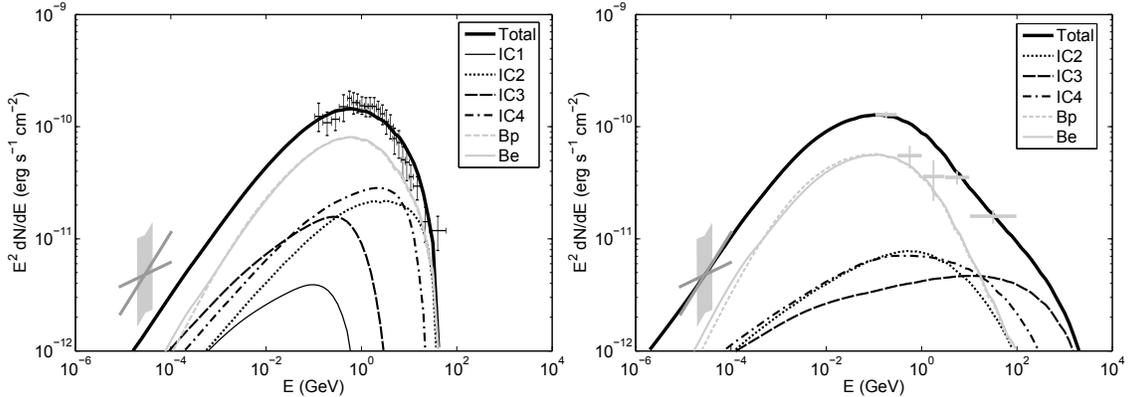}
\end{center}
\caption{Spectrum of different components of the emission expected in leptonic models. Here IC1 corresponds to
inverse-Compton scattering on IR photons produced by CND, IC2 - on UV photons produced in the central parsec,
IC3 - on IR photons produced in the central parsec, and IC4 - on visible photons produced by NSC;
Bp corresponds to electron-ion bremsstrahlung and Be - to electron-electron bremsstrahlung.
Black crosses represent data from \citet{masha15} and gray crosses data from 3FGL.
Gray area shows the hard X-ray flux observed by \nustar~ with X-shaped mark demonstrating the allowed minimum and maximum spectral indices.
{\it Left:} Injection index is $\alpha = 2.1$. {\it Right:} Injection index is $\alpha = 2.5$ (for 3FGL data).
}
\label{fig:leptonic_sed}
\end{figure}

To estimate the potential effect of a spectral break in the injected spectrum of electrons
we introduced a low-energy cut-off and recalculated the intensity of hard X-ray emission.
We took injection spectrum with index of $\alpha = 2.5$ as it should be the most significantly affected one.
Positions of the cut-off were chosen to be at $E_{min} = 1$ MeV, 10 MeV, 30 MeV and 100 MeV. Gamma-ray data do not allow us to place a cut-off above
100 MeV. The corresponding hard X-ray flux in these cases are 1.0, 0.7, 0.4 and 0.2 of the original value. In more realistic scenarios with a
spectral break instead of a low-energy cut-off the expected drop of the X-ray emission should be lower.
Therefore we consider a single power-law spectrum is a reasonable assumption taking into account
that actual X-ray emission may be reduced by a factor of few.

Different components of the emission contributing to X-rays and gamma-rays are presented in Figure~\ref{fig:leptonic_sed}.
One can see that emission is clearly dominated by bremsstrahlung process.
Therefore, the intensity of hard X-ray emission may be higher if we underestimated the ratio
between soft photons and gas in the central few parsecs.

\begin{figure}[h]
\begin{center}
\includegraphics[width=0.9\textwidth]{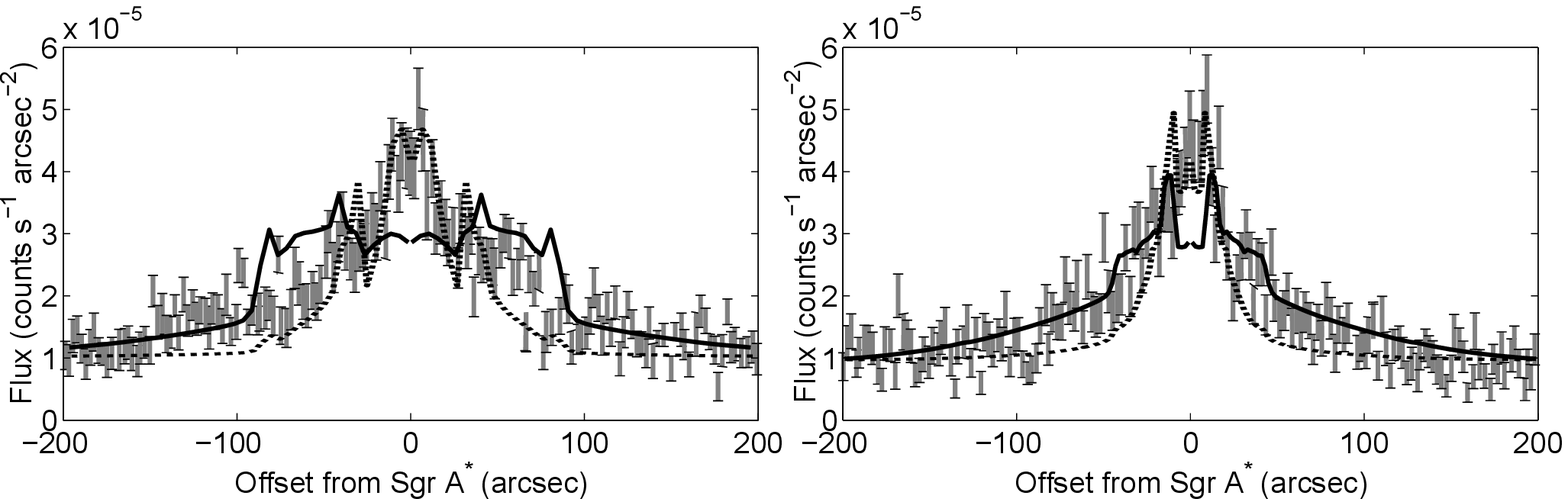}
\end{center}
\caption{
Spatial profile of the hard X-ray emission along the Galactic longitude (on the left)
and latitude (on the right) in the case of leptonic models. Solid curves correspond to single-burst model
at $t = 700$ years and dotted curves to continuous supply model. Data are taken from \citet{perez15}.
}
\label{fig:leptonic_spat}
\end{figure}

Spatial profile of the hard X-ray emission is also reproduced quite nicely (see Figure~\ref{fig:leptonic_spat}).
To compare the \nustar~ data taken from \citet{perez15}, we multiply the computed X-ray flux by some number
to match the data and added a constant background flux of the order of $10^{-5}$ counts s$^{-1}$ arcsec$^{-1}$.
Excess emission observed at the position of Sgr A$^*$ may be due to X-ray filament G359.97$-$0.038.

\begin{figure}[h]
\begin{center}
\includegraphics[width=0.9\textwidth]{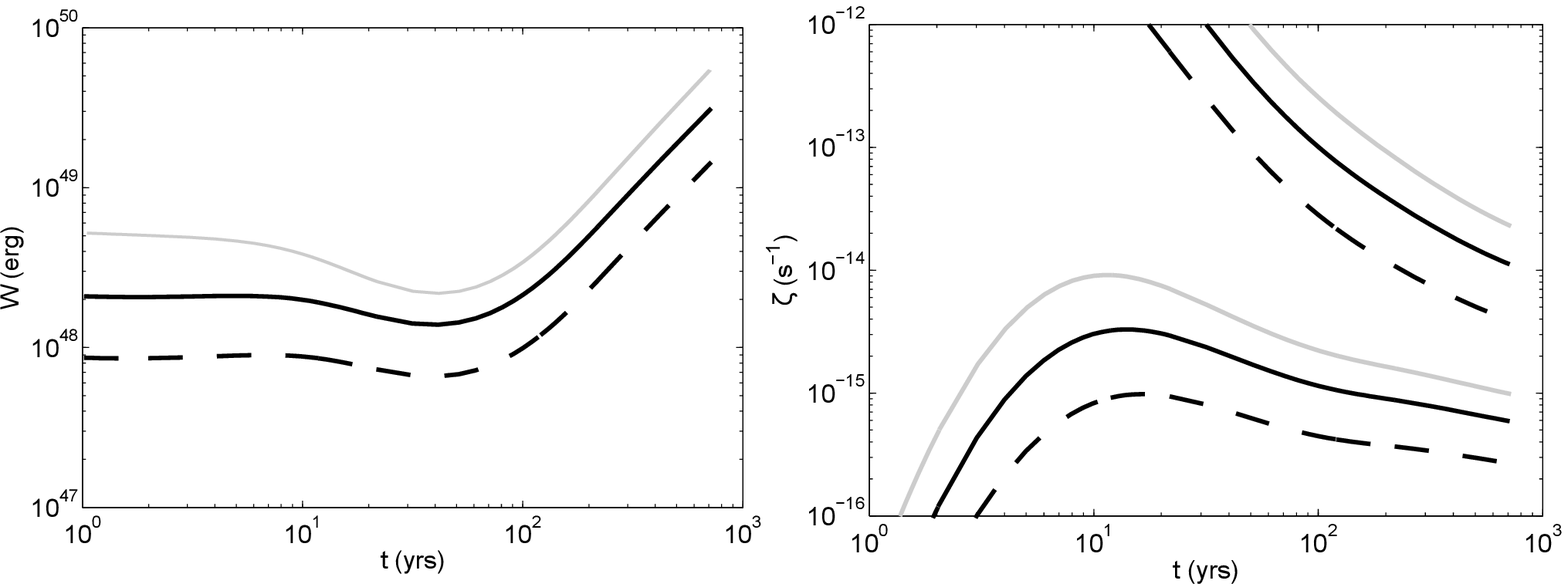}
\end{center}
\caption{{\it Left:} Total energy of injected electrons required to reproduce observed gamma-ray emission
in the case of single-burst leptonic model
depending on how long ago the injection of the particles took place.
Black solid line is the case of injection index $\alpha = 2.3$,
black dashed line is $\alpha = 2.1$, and gray line is $\alpha = 2.5$.
{\it Right:} Ionization rate in the case of single-burst leptonic model.
Top curves - in the central cavity, bottom curves - averaged over CND.
Other notations are the same.
}
\label{fig:leptonic_W}
\end{figure}

The total energy budget required to produce observed gamma-ray emission depends on how long ago the injection took place
(see the left panel of Figure~\ref{fig:leptonic_W}). Since electrons are affected by energy losses and escape process,
it is natural that the required energy increases as the time from the injection increases.
One can see that the overall
energy is quite high. However in all cases, injection can be produced by a single supernova explosion with electron
acceleration efficiency of about 10\%.

In the right panel of Figure~\ref{fig:leptonic_W}, we show the evolution of ionization rate of molecular hydrogen
estimated in two locations, (1) in the central cavity, and (2) averaged over CND.
One can see that in both cases at $t = 7\times 10^2$ years the estimated values do not contradict the measured value
$\zeta \simeq 1.2\times 10^{-15}$ s$^{-1}$ \citep{goto13,goto14}.
Shape of the injected spectrum at low energies have higher impact on
the ionization rate in comparison to the impact on the X-ray flux. For spectra with cut-off
at $E_{min} = 1$ MeV, 10 MeV, 30 MeV and 100 MeV, the value of ionization rate in the central cavity becomes
1.0, 0.45, 0.12, 0.03 of the original value. The corresponding values of ionization rate averaged over CND
are 1.0, 0.75, 0.4, 0.16 of the original value and they are very close to the corresponding modifications of the X-ray flux.
Therefore ionization rate in the central cavity can be slightly reduced by an adjustment of the spectral shape.
On the other hand the value of ionization rate averaged over CND is tied to the intensity of the hard X-rays and
there are not much freedom. In any case, as we mentioned before the measured ionization rate is uncertain
up to an order of magnitude in values and cannot be considered as a very reliable restriction.

Another important indicator related to ionization is the intensity of 6.4 keV K$\alpha$ line of neutral iron.
For solar abundance of iron we obtained the following intensities of the emission from CND,
(1) for $\alpha = 2.1$ we expect $I_{6.4} = 1.3 \times 10^{-7}$ ph cm$^2$ s$^{-1}$,
(2) for $\alpha = 2.3$ we expect $I_{6.4} = 2.3 \times 10^{-7}$ ph cm$^2$ s$^{-1}$, and
(3) for $\alpha = 2.5$ we expect $I_{6.4} = 2.8 \times 10^{-7}$ ph cm$^2$ s$^{-1}$.
It is worth noting that if we consider the whole region of interest, then the expected equivalent width of this line will be about 15 eV.
This value is extremely low since only a small part of X-ray emission comes from the molecular gas.
Moreover, since this region is also contaminated by thermal emission from Sgr A East which we did not take into account.
One can expect even lower values of the equivalent width of this line.

\begin{figure}[h]
\begin{center}
\includegraphics[width=0.4\textwidth]{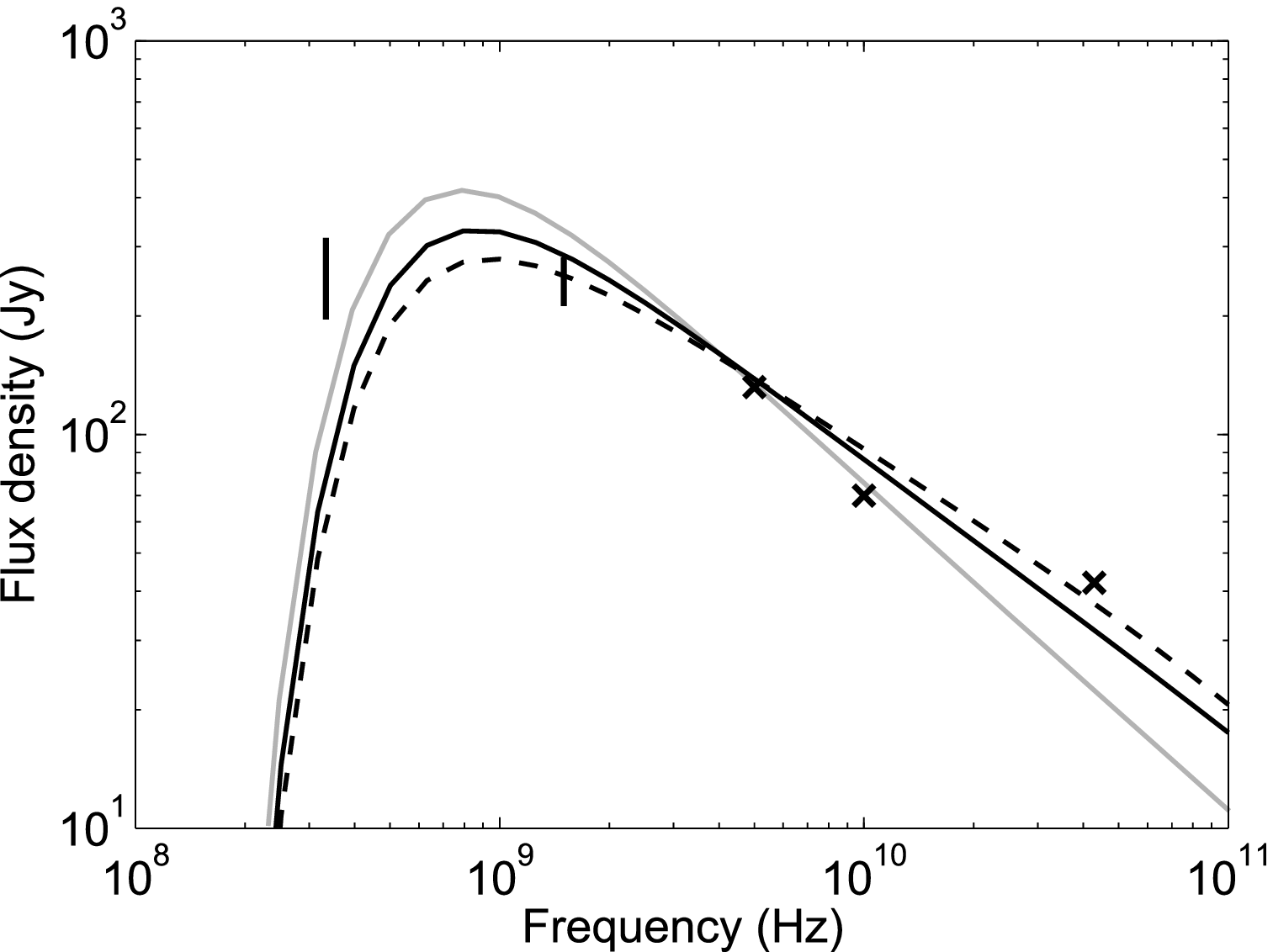}
\end{center}
\caption{Spectrum of the radio emission from the radio halo in the case of single-burst leptonic model.
Black solid line is the case of injection index $\alpha = 2.3$, black dashed line is $\alpha = 2.1$,
and gray line is $\alpha = 2.5$.
Data are taken from \citet{pedlar89}.
}
\label{fig:leptonic_radio}
\end{figure}

The expected radio emission from radio halo is shown in Figure~\ref{fig:leptonic_radio}.
Experimental data were taken from \citet{pedlar89}.
To match the data-points, we assumed that the average magnetic field is of the order of $H= 44$ $\mu$G in the case of $\alpha=2.1\sim 2.3$,
and $H= 54$ $\mu$G in the case of $\alpha=2.5$. We also take into account of free-free absorption by the plasma of radio halo.
One can see that the data point at low frequency is fitted poorly, but we can assume that there is an additional thermal component
of radio emission as assumed by \citet{pedlar89}.

\subsection{Continuous supply leptonic model}\label{sec:cslm}
This model assumes that central source supplies energetic electrons continuously.
The temporal term of the injection function in Equation~(\ref{eq:eq_Q_genform_e}) is expressed as $T(t) = \Theta(t)$.
For the energy part we assume that the spectral index is $\alpha = 2.5$ and the cut-off energy is $E_{\rm max} = \infty$.

\begin{figure}[h]
\begin{center}
\includegraphics[width=0.9\textwidth]{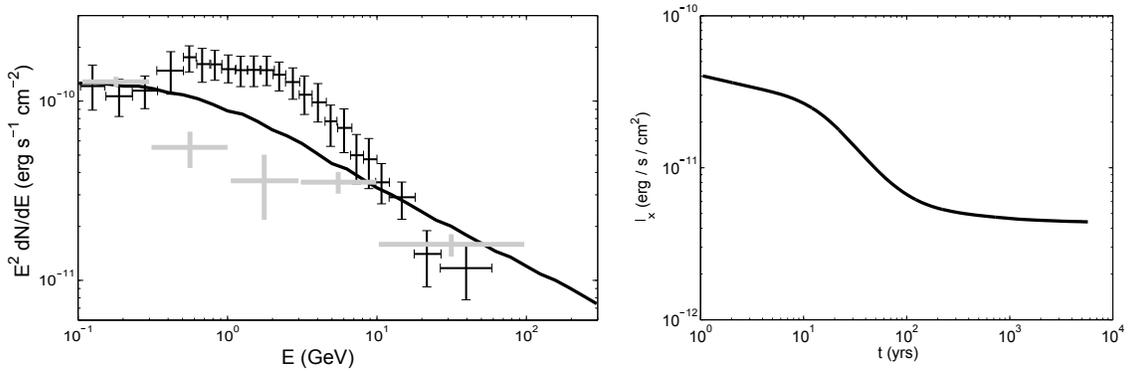}
\end{center}
\caption{The same as Figure~\ref{fig:hadronic} except for the case of continuous supply leptonic model.}
\label{fig:leptonic_stat}
\end{figure}

Gamma-ray emission and evolution of hard X-ray flux from the model looks promising as shown in Figure~\ref{fig:leptonic_stat}.
The spatial distribution of X-ray emission is slightly narrower than the one observed by NuSTAR,
see dotted line in Figure~\ref{fig:leptonic_spat}.
Nevertheless, it is still consistent with the observational data.
Total energy required to reproduce the data is about $4.6\times 10^{38}$ erg s$^{-1}$.
The only problem of the continuous supply model is that the ionization rate $\zeta$ in the central cavity exceeds $10^{-11}$ s$^{-1}$
which is more than 4 orders of magnitude higher than reported by \citet{goto14}.
On the other hand, average ionization rate inside the CND is about $3\times 10^{-15}$ s$^{-1}$
and does not differ significantly from the experimental value.

The reason for the very high ionization rate in the central area originate from the characteristic of the source of particles there.
Indeed, stationary solution gives a $1/r$ distribution near the center, which produces a huge spike of particle density in the
central region.
Now suppose we turn off the source and wait for about 100 years, the particles will spread out and the ionization rate will drop significantly
to a more reasonable value of about $\zeta\simeq 5\times 10^{-14}$ s$^{-1}$.

Another way to reduce the ionization rate is to assume that injected spectrum has a spectral break at low energies.
Despite the fact that such modification does not affect the average particle distribution at large distances
where it is modified by energy losses the low-energy spectral shape may be important near the source.
This assumption may decrease the expected ionization rate near the center even further.

\section{Mixed models}\label{sec:mixed}
\begin{figure}[h]
\begin{center}
\includegraphics[width=0.6\textwidth]{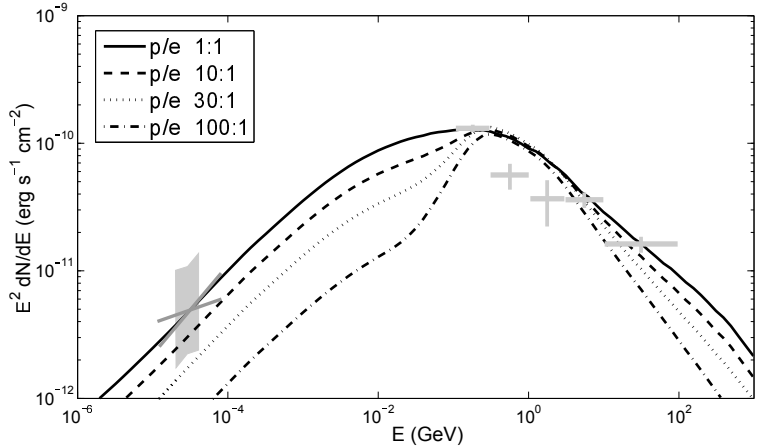}
\end{center}
\caption{Multi-wavelength spectrum of the emission expected in the mixed models for different
proton-to-electron density ratio above 1 GeV. The injection particles spectral index is $\alpha = 2.5$.
Gray crosses represent gamma-ray data from 3FGL \citep{3fgl}.
Gray area represent data from \nustar~ and the X-shaped mark shows the allowed minimum and maximum spectral indices.
}
\label{fig:mixed_sed}
\end{figure}

It is more common and conceivable that both primary electrons and protons are accelerated in the same astrophysical source.
For example, the observed composition of cosmic rays in the vicinity of the Earth suggests that potential sources of
cosmic rays produce protons and electrons above 1 GeV in a ratio from 100:1 to 50:1 \citep{strong10}.
To take this possibility into account, we made a linear combination of hadronic and leptonic models.
We assumed that both primary protons and primary electron have the same injection spectral index $\alpha = 2.5$.
The resulting spectrum is shown in Figure~\ref{fig:mixed_sed}.
One can see that for near Earth proton-to-electron ratio ($100:1$) the hard X-ray flux is too low because the emission
is dominated by the hadronic component.
However, if we reduce the proton-to-electron ratio to 30:1, charged particles can make a non-negligible contribution to the hard X-ray flux.

It is worth noting that proton-to-electron ratio deduced here is much higher than the one deduced from direct
comparison of energy budgets from Sections~\ref{sec:hadronic} and \ref{sec:leptonic}. The reason for this discrepancy is that here we use
propagation parameters from Section~\ref{sec:leptonic} for both hadronic and leptonic components. These parameters significantly reduce gamma-ray
emission from protons in comparison with those from Section~\ref{sec:hadronic}. Therefore proton-to-electron ratio becomes higher.

\section{Conclusions}\label{conclusions}
We analyzed the propagation of the charged particles emitted by a central source in the GC and their interactions with the surrounding medium
and radiation field. Our goal is to explain the observed multi-wavelength emissions from the inner few parsecs of the Galaxy.
We considered different scenarios: the injected particles can be either protons or electrons (or mixed)
and the source may be transient or continuous.

Our analysis shows that pure hadronic injection model is not good enough to explain both X-ray and gamma-ray observations.
If the hadronic model is able to reproduce the observed gamma-ray intensity, the hard X-ray flux is more than an order of
magnitude lower than the observed one.
We note, however, that the X-ray flux depends strongly on the density of soft photons in the region.
If for some reason the density of IR photons is an order of magnitude higher than what we have estimated,
the expected X-ray flux could be comparable with observations.
Despite we do not know what kind of sources may increase IR photon density by an order of magnitude, such a high density
could be justified if the cut-off in the TeV spectrum of the central source is indeed due to absorption of the IR photons.
Moreover, a non-negligible contribution from synchrotron emission to the hard X-rays may be expected in this case.

On the other hand, pure leptonic injection models are able to provide reasonable X-ray fluxes.
Electrons are also able to reproduce gamma-ray spectrum observed by \fermi~ better than hadronic models \citep[see, e.g.,][]{masha15}.
Both transient and continuous models can reproduce the spatial and spectral shape of the hard X-ray emission.
The only shortcoming of the continuous model is that the ionization rate in the central cavity exceeds the observed one
by 4 orders of magnitude, although IR estimation of the ionization rate may not be very reliable.
The problem of exceedingly high ionization rate can be resolved if we assume that the continuous source
was somehow inactive for the last 100 yrs.

It is interesting to note that similar conclusion was made from X-ray observations of molecular clouds in the GC
\citep[see][and references therein]{inu09,ponti10,terrier,nob11}.
It was concluded that just about 100 years ago Sgr A$^*$ was a source of hard X-rays with a luminosity $\sim 10^{39}$ erg s$^{-1}$.
This number almost exactly coincides with our findings.
Even if this is just a coincidence, it would be interesting to find a possible connection between these phenomena.

In the mixed model that both protons and electrons are emitted by the central source,
relatively low proton-to-electron ratio is necessary.
The ratio of the density of proton above 1 GeV to the density of electron above 1 GeV must be smaller than 20.
For comparison this proton-to-electron ratio at Earth is around $50\sim 100$.
Nevertheless, we should point out that this inferred ratio depends on ratio between gas density and soft photons density.
If the density of radiation field is underestimated or if GeV particles are unable to penetrate inside the molecular cloud,
then even at a higher proton-to-electron ratio the model could be tuned to reproduce the observed X-ray flux.

\acknowledgments
DOC and VAD acknowledge support from the RFFI grants 15-52-52004 and 15-02-02358.
KSC is supported by the GRF Grants of the Government of the Hong Kong SAR under HKU 17310916.
DOC, KSC, and VAD acknowledge support from the International Space Science Institute to the International Team
``New Approach to Active Processes in Central Regions of Galaxies''.
CMK is supported, in part, by the Taiwan Ministry of Science and Technology grants MOST 104-2923-M-008-001-MY3
and MOST 105-2112-M-008-011-MY3.
The authors are grateful to Charles Hailey for very useful comments about \nustar~ data and their interpretation.

\appendix
\section{Soft photon density in the Galactic Central Region}\label{appx:wop}
Our approach to calculate the density of soft photons in the Galactic center (GC) is similar to the one used by \citet{kist15}.
We split the overall soft photon field into four different components, each with its own temperature and spatial distribution.
Explicitly, we consider UV and IR emission from the central parsec,
IR emission from the circumnuclear disk (CND) and optical emission from the Nuclear Star Cluster (NSC).

In the case of the emissions from the central parsec we adopt the data from \citet{mezger}.
According to the paper the temperature and total luminosity of the UV emission are $T^{\rm UV} = 3\times 10^4$ K and
$L_{\rm UV}=(7.5 \pm 3.5) \times 10^7$ L$_\odot$, respectively.
The corresponding average energy density of UV photons inside the central parsec is $(5.5\pm 2.5)\times 10^4$ eV cm$^{-3}$.
We take the lowest possible estimate and assume that the average energy density of UV photons is
$w_0^{\rm UV} = 3\times 10^4$ eV cm$^{-3}$.
As for IR emission, we adopt $T^{\rm IR} = 170$ K and $w_0^{\rm IR} = 3.5\times 10^3$ eV cm$^{-3}$.

Outside the central parsec we assume that absorption is not important and the distribution of soft photons evolve as
\begin{equation}
w(R) = \left\{
\begin{array}{ll}
w_0 &\mbox{, if R $<$ 1 pc} \\
w_0\left(\frac{R}{1~\mbox{pc}}\right)^{-2} & \mbox{, if R $\geq$ 1 pc, }
\end{array}
\right.
\end{equation}
where $R^2 = r^2 + z^2$.

The last two components do not have a spherically symmetric distribution and their distribution should be estimated based on
the distribution of sources. If the emissivity of the sources is $\epsilon(r^\prime,z^\prime)$,
the density of soft photons can be estimated as (ignore absorption)
\begin{equation}\label{eq:eps_to_w}
w(r,z) = \frac{1}{4\pi c}\int r^\prime \,dr^\prime \int dz^\prime
\int \frac{\epsilon(r^\prime,z^\prime)\,d\phi^\prime}{\left[ (z-z^\prime)^2 + r^2 + r^{\prime\,2} - 2rr^\prime\cos \phi^\prime \right]}\,.
\end{equation}

In the case of IR emission from the CND, we assume that the density of sources is proportional to the density of gas described by
\citet{katia12}. According to \citet{mezger} and \citet{telesco} the temperature of the emission is $T^{\rm CND} = 70$ K
and the total luminosity is
\begin{equation}\label{eq:CND}
L^{\rm CND} = \int 2\pi r^\prime\, dr^\prime \int \epsilon^{\rm CND}(r^\prime,z^\prime)\,dz^\prime = 10^{40} {\rm erg\ s}^{-1}\,.
\end{equation}

In the case of the emission from the NSC we need to estimate the density of sources first.
According to \citet{schodel} the surface brightness of NSC can be described by a S\'{e}rsic profile
\begin{equation}\label{eq:Sersic}
I(x,y) = I_{\rm e} \exp\left(-b_n\left[\left(\frac{\varrho}{R_{\rm e}}\right)^{1/n} - 1\right]\right)\,,
\quad
\varrho^2 = x^2 + \frac{y^2}{q^2}\,,
\end{equation}
where $\varrho$ is called the modified projected radius, and $q = 0.7$ is the projected minor to major axis ratio.
Here $x$ and $y$ are the vertical and horizontal coordinates and $n$, $b_n$, $R_{\rm e}$ and $I_{\rm e}$ are fitting parameters.
Their numerical values can be found in Table~3 of \citet{schodel}.
We note that specific choice of the model from that table only slightly affects the final result.

Following \citet{schodel} we assume that the distribution of sources depends only on the modified radius $\rho$
\begin{equation}\label{eq:NSC}
\epsilon^{\rm NSC}(r,z) = \epsilon(\rho)\,,
\quad
\rho^2 = r^2 + \frac{z^2}{q_s^2} \,,
\end{equation}
where $q_s$ is the true minor to major axis ratio. Generally speaking $q_s \neq q$.
To simplify and to keep the symmetry of our model, we assume that the plane of symmetry of NSC is parallel to that of CND,
i.e., inclined by 20$\degr$ to the Galactic plane. The inclination makes projected minor to major axis ratio $q$
slightly larger than the real one $q_s$.

Integrating the emissivity along the line-of-sight gives the surface brightness
\begin{equation}\label{eq:surface_brt}
I(x,y) = \frac{1}{4\pi}\int\limits_{-\infty}^\infty \epsilon^{\rm NSC}(r,z)\,d\ell\,,
\end{equation}
where the inclination of $\theta = 20\degr$ is taken into account,
\begin{equation}\label{eq:inclination}
r^2 = x^2 + \ell^2 \cos^2 \theta\,,
\quad
z = y + \ell\sin \theta\,.
\end{equation}

We transform Equation~(\ref{eq:surface_brt}) into discrete form and solve the resulting system of linear equations
to obtain the source distribution
\begin{equation}\label{eq:discrete}
I_j = A_{i,j}\epsilon_j \,,
\end{equation}
where $A_{i,j}$ is the matrix produced by applying Simpson's numerical integration rule to Equation~(\ref{eq:surface_brt}).
The value of $q_s$ can be estimated: for apparent ratio $q = 0.7$ the real ratio is $q_s = 0.6$.

Knowing the source distribution $\epsilon^{\rm NSC}(r,z)$, we apply Equation(\ref{eq:eps_to_w}) to obtain the distribution of
soft photons $w^{\rm NSC}(r,z)$ produced by NSC. We assume that the temperature of this emission is 3200 K.


\end{document}